\documentclass[CEJP,DVI]{cej} 

\newcommand{\be}{\begin{eqnarray}}
\newcommand{\ee}{\end{eqnarray}}

\newcommand{\hmu}{\hat{\mu}}

\makeatletter
\newcommand{\fslash}[2][0mu]{%
    \mathchoice
     {\fsl@sh\displaystyle{#1}{#2}}%
     {\fsl@sh\textstyle{#1}{#2}}%
     {\fsl@sh\scriptstyle{#1}{#2}}%
      {\fsl@sh\scriptscriptstyle{#1}{#2}}}
    \newcommand{\fsl@sh}[3]{%
    \m@th\ooalign{$\hfil#1\mkern#2/\hfil$\crcr$#1#3$}}
\def\lsim{\raise0.3ex\hbox{$<$\kern-0.75em\raise-1.1ex\hbox{$\sim$}}}
\def\gsim{\raise0.3ex\hbox{$>$\kern-0.75em\raise-1.1ex\hbox{$\sim$}}}

\usepackage{layout}
\usepackage{amsmath}
\usepackage{textcomp}
\usepackage{hyperref}

\title{Probing QCD chiral cross over transition in heavy ion collisions with fluctuations}

\articletype{ Communication}

\author{Krzysztof Redlich\email{redlich@ift.uni.wroc.pl}}

\institute{
Institute of Theoretical Physics, University of Wroc\l aw, PL--50204
  Wroc\l aw, Poland
     \hskip 8.5cm
ExtreMe Matter Institute EMMI, GSI, D-64291 Darmstadt, Germany}
\abstract{
 We argue that by measuring higher moments of the net proton number fluctuations in heavy ion collisions  (HIC) one can probe the QCD chiral cross over
  transition experimentally.
 We discuss the properties of fluctuations of the net baryon number  in the vicinity of the chiral crossover transition within  the Polyakov loop
 extended quark-meson model at finite temperature and  baryon density. The calculation includes non-perturbative dynamics implemented within the
  functional renormalization group approach.  We find a clear signal for the chiral crossover transition in the fluctuations of the net baryon number.
   We address our theoretical findings to experimental data of STAR Collaboration on energy and centrality dependence
   of the net proton number fluctuations and their probability distributions in HIC.
}

\keywords{QCD phase transition and phase diagram
\*\ Chiral symmetry and charge fluctuations
\*\ Heavy ion collisions }

\begin{document}
\maketitle

\section{QCD phase diagram}
Although, the important  question addressed in QCD on the existence of  a true 2nd order phase transition  at finite chemical potential (critical point)
 has    not been answered yet, nevertheless,  there is recently an essential progress in the  quantitative description of the QCD phase diagram.
The  lattice QCD (LQCD)  provided a final   value for the chiral cross over transition temperature at vanishing chemical potential \cite{l1}. The LQCD
has also provided arguments that at small $\mu\simeq 0$  the chiral cross over line  is  the pseudo critical line of the    2nd order
 chiral phase transition belonging  to the universality class
of 3-dimensional, O(4) symmetric spin models \cite{l2}. Based on the universality arguments the LQCD has also provided
the curvature of the chiral transition  line, $T_c(\mu_c)$ \cite{l2}. These results confirm that,  at least at small values of the  baryon chemical potential $\mu$,
the chiral cross over transition appears in the near vicinity to the chemical freezeout line  \cite{f1} obtained from the analysis of particle yields 
 measured in HIC.

  The numerical coincidence of thermal parameters  for  the chiral transition and the freezeout conditions indicates that the hadron resonance gas partition
   function, which describes  chemical equilibration of particle yields in HIC,   should  describe also the QCD thermodynamics up to a near
    vicinity to  the transition to a quark gluon plasma phase. Indeed,  the equation of state calculated on the lattice  as well as other thermodynamical
     observables    in the hadronic phase,  were shown to be very well quantified by the hadron resonance gas (HRG) partition function \cite{h1}.

Already at vanishing  chemical potential, {\it i.e.}
under conditions realized in the high energy runs at RHIC or LHC, the
question  arises to what extent a refined analysis of freeze-out
conditions can establish the existence of a chiral phase transition.
In this lecture we will argue that even  at
 $\mu_B/T \simeq 0$ the net baryon  number fluctuations and their higher moments
can be used to  identify the chiral cross over transition experimentally \cite{o1,o2,o3,o4}.
\begin{figure}[t]
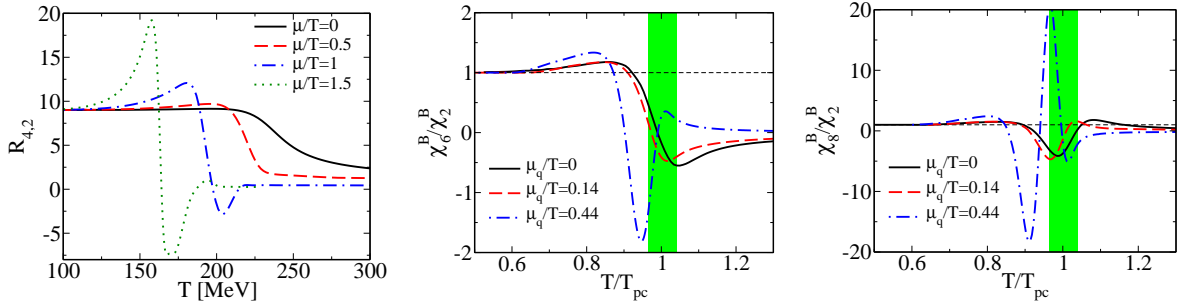

\begin{center}
\hspace*{-0.5cm}
\includegraphics*[width=4.99cm,]{R_frg.eps}\hspace*{0.3cm}
\includegraphics*[width=4.7cm,]{c6c2mu02.eps}\hspace*{0.3cm}
\includegraphics*[width=4.9cm,]{c8c2mu02.eps}
\vspace*{-0.5cm}\vskip 0.0cm\caption
{
The temperature dependence of kurtosis $R_{4,2}:= 9\chi_{4}^B$/$\chi_{2}^B$ and the higher order,   $\chi_{6}^B$/$\chi_{2}^B$ and $\chi_{8}^B/\chi_{2}^B$,
 ratios of cumulants  for different  $\mu_q/T$ calculated in the PQM   model within the FRG  approach
  \cite{o2,o4}.
The $T_{pc}$  is the pseudo-critical temperature obtained in the model at the physical pion mass.  }
\label{fig:PD}
\end{center}
\end{figure}

\section{Charge fluctuations and the chiral cross over transition}

Due to   remnants of O(4) criticality related with the chiral phase transition observed in LQCD,  the free energy ($f$)
 near   the chiral phase
transition temperature $T_c$
may be represented  in terms of
  singular ($f_s$) and regular contributions ($f_r$) as
\begin{equation}
f(T,\mu_q,m_q) = f_s(T,\mu_q,m_q) + f_r(T,\mu_q,m_q) \; ,
\label{freeenergy}
\end{equation}
where in addition  to the temperature $T$ we also introduced explicit
dependence on the light quark chemical potential, $\mu_q=\mu_B/3$,
and the (degenerate) light quark masses $m_q\equiv m_u = m_d$.
The singular part of the free energy may be written as \cite{s1}
\begin{equation}
f_s(T,\mu_q,h) = h^{1+1/\delta} f_s(z) \ ,\ z \equiv t/h^{1/\beta\delta}
\label{singular}
\end{equation}
with $\beta,\ \delta$ are critical exponents of the 3-dimensional,
$O(4)$ universality class  and
$t \equiv \frac{1}{t_0}\left( \frac{T-T_c}{T_c} +
\kappa_q \left( \frac{\mu_q}{T}\right)^2
\right)
$,
$
h \equiv \frac{1}{h_0} \frac{m_q}{T_c} $.
Here $T_c$ is the phase transition temperature in the chiral limit
and $t_0$, $h_0$ are non-universal scale parameters.
 The proportionality constant
$\kappa_q\simeq 0.06$,
 has recently been determined from
a scaling analysis in (2+1)-flavor QCD \cite{l2}.
The scaling function $f_s$  and its derivatives  have recently been
calculated   using high precision Monte Carlo simulations of the
3-dimensional O(4) spin model
\cite{e1}.

We want to focus here on properties of moments of net baryon number
fluctuations, which are obtained from Eq.~\ref{freeenergy} by taking
derivatives with respect to $\hmu_B = \mu_B/T$,
\begin{equation}
\chi_{n}^{B} = - \frac{1}{3^{n}}
\frac{\partial^{n}f/T^4}{\partial\hmu_q^n}= - \frac{1}{3^{n}}
\frac{\partial^{n}f_r/T^4}{\partial\hmu_q^n} - \frac{1}{3^{n}}
\frac{\partial^{n}f_s/T^4}{\partial\hmu_q^n} = \chi_{n,r}^{B}+\chi_{n,s}^{B} \; .
\label{obs}
\end{equation}
In the hadronic  phase and away from transition temperature the regular part $\chi_{n,r}^{B}$ should be well described
 by the hadron
resonance gas partition function which will be a reference for critical fluctuations coming from the singular part $\chi_{n,s}^{B}$. Thus,
 any deviations from the regular i.e.  hadronic gas contribution  could be an indication of criticality due to  remnants of the  chiral O(4) transition.

Higher order moments will become increasingly sensitive to the
singular part of the free energy. From Eq.~\ref{singular}
it is apparent that these moments show a strong quark mass dependence
in the vicinity of the critical temperature,
\begin{equation}
\chi_{n,s}^B \sim
\begin{cases}
-(2\kappa_q)^{n/2} h^{(2-\alpha -n/2)/\beta\delta} f_s^{(n/2)}(z)
& ,\ {\rm for}\ \mu_q /T = 0,\
{\rm and}\ n\ {\rm even} \\
- (2\kappa_q)^n \left( \frac{\mu_q}{T} \right)^n
h^{(2-\alpha -n)/\beta\delta} f_s^{(n)}(z)
&,\ {\rm for}\ \mu_q/T > 0
\end{cases}\ .
\label{fluct_mass}
\end{equation}
where we used $2-\alpha = \beta\delta (1+1/\delta)$.
As $\alpha = -0.2131 (34)$ is negative in the 3-dimensional, $O(4)$ universality class,
the 4th  order moments of the net baryon number fluctuations
do not  diverge yet in the chiral limit at the chiral transition
temperature, $z=0$. The first divergent moment is obtained
for $n= 6$ if $\mu_q/T=0$ and for $n= 3$ if $\mu_q/T > 0$.

\begin{figure}[t]
\begin{center}
\hspace*{1.5cm}
\includegraphics*[width=5.cm,height=4.3cm]{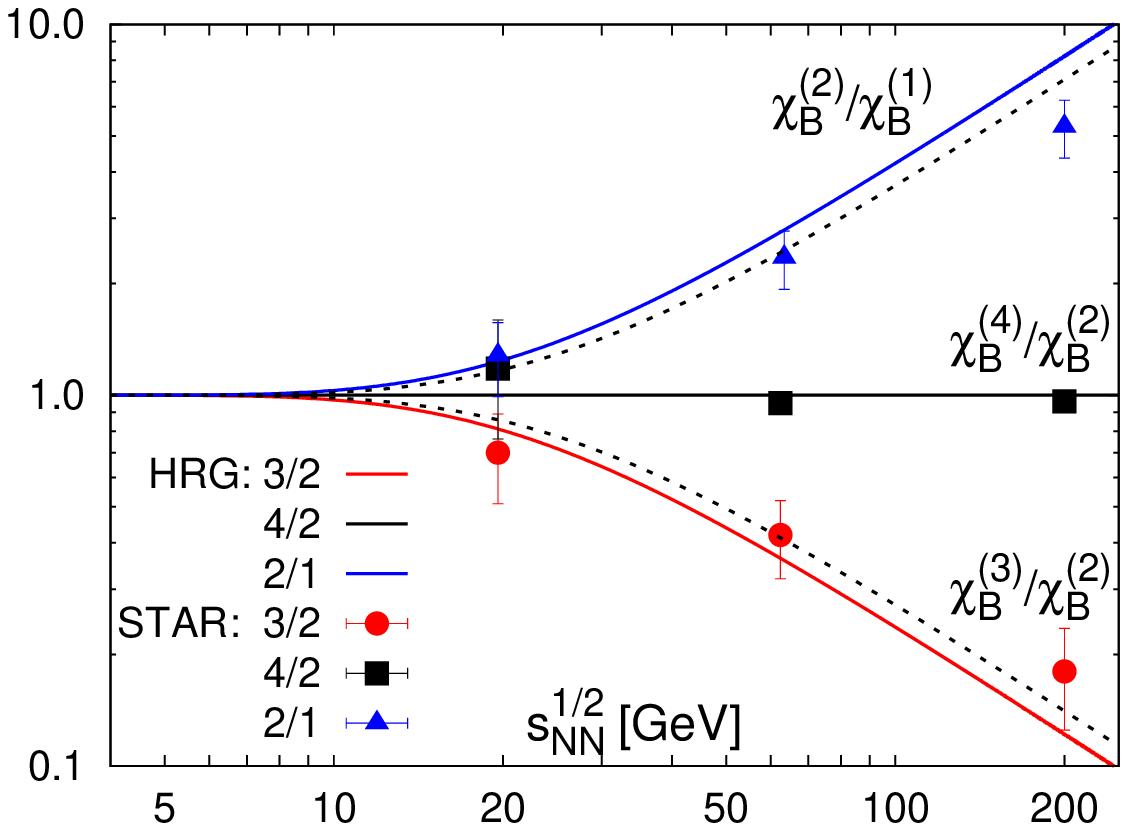}\hspace*{0.5cm}
\includegraphics*[width=3.7cm,height=4.7cm]{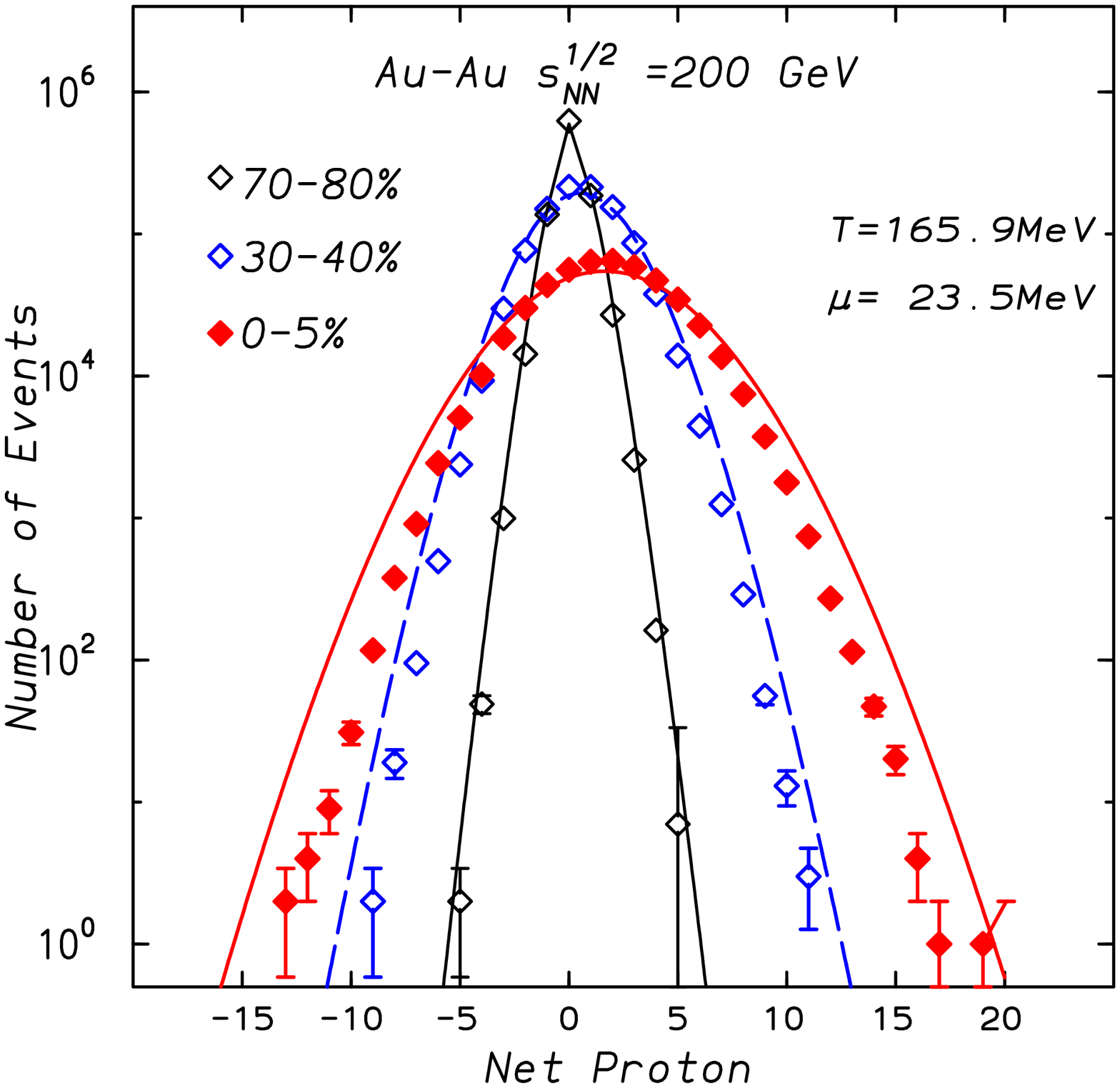}\hspace*{0.5cm}
\includegraphics*[width=4.1cm,height=4.9cm]{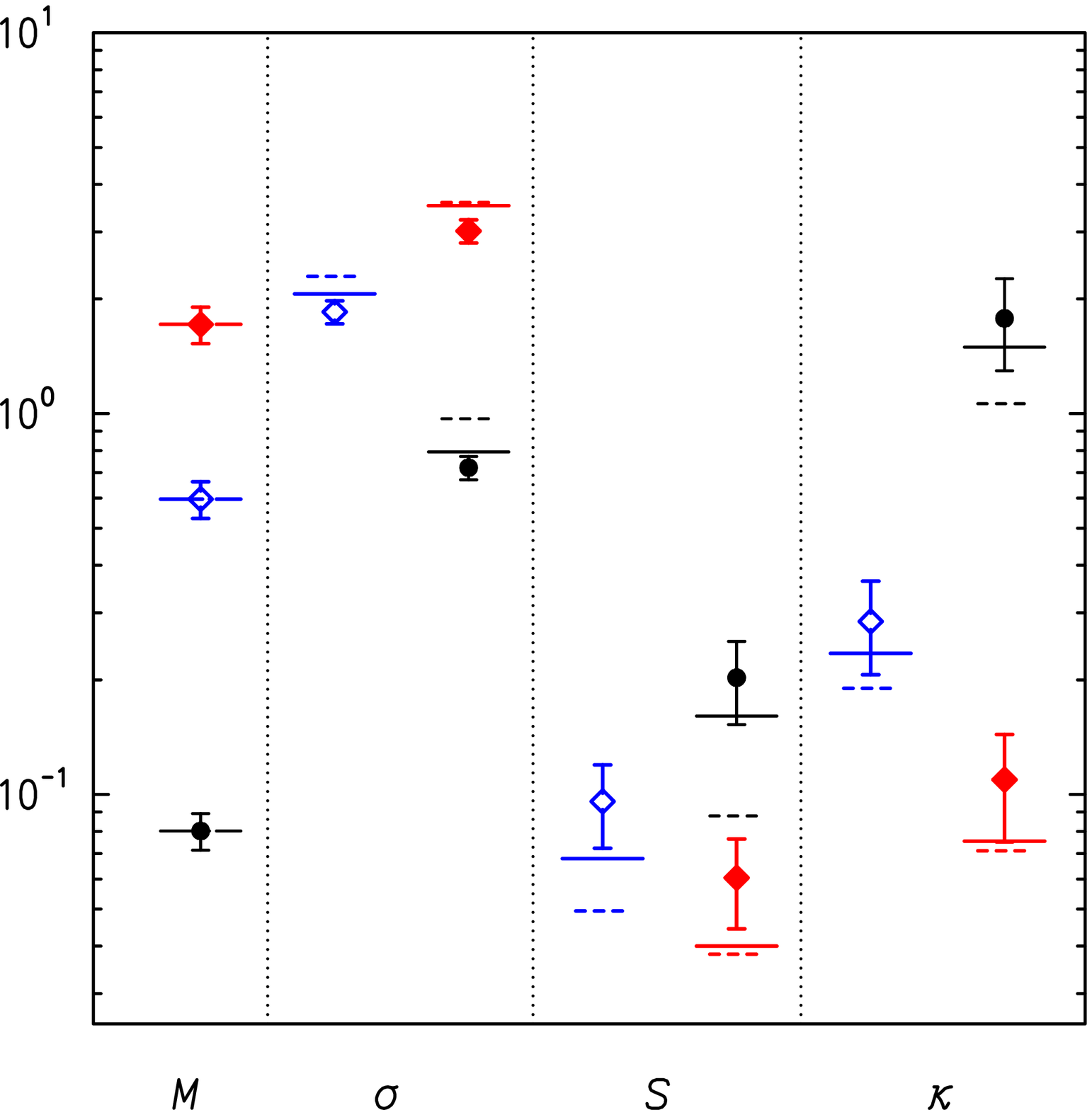}\hspace*{0.5cm}
\vspace*{-0.5cm}
\vskip 0.0cm\caption
{
Left-hand figure: the ratio of quadratic fluctuations and mean net baryon number ($\sigma^2/M$),
cubic to quadratic  ($S\sigma$) and
quartic to quadratic ($\kappa\sigma^2$) baryon number fluctuations
calculated in the HRG model on the freeze-out curve \cite{o1} and compared to results
obtained by the STAR Collaboration \cite{star}.
The dashed curves show the approximate $\tanh (\mu_B/T)$
result for $\kappa\sigma^2$ and $S\sigma$, respectively. Middle figure: The probability distributions, uncorrected for
 event-by-event counting efficiency,
 for the net proton number
for different centralities taken by STAR Collaboration in Au-Au collisions at $\sqrt {s_{NN}}=200$ GeV \cite{star}. The lines are the Skellam distributions
calculated within HRG model \cite{o3}. Right-hand figure: Mean (M), variance ($\sigma$),
skewness (S) and kurtosis ($\kappa$) of the net proton number calculated from the probability distributions shown in the middle figure.
}
\label{fig:PD}
\end{center}
\end{figure}

In the hadron resonance gas (HRG) the regular part of the fluctuations $\chi_{n,r}^{B}$ can be directly calculated from the
thermodynamic pressure following Eq. {\ref{obs}}. The HRG  is a mixture of ideal gases of all particles  and resonances, consequently
 the thermodynamic pressure exhibits a factorization of $T$ and $\mu_B/T$ dependence. Under the Boltzmann approximation the pressure in
 the HRG,  $P^{HRG}(T,\mu_B)\simeq f(T)\cosh (\mu_B/T)$, where $f(T)$
  contains  contributions from all baryons and baryonic  resonances. With such $P^{HRG}$  there  are  particular
  properties of  ratios of cumulants,
\begin{equation}
{{\chi_{2n,r}^B}\over {\chi_{2,r}^B}}|_{HRG}=1~~,~~{{\chi_{(2n+1),r}^B}\over {\chi_{1,r}^B}}|_{HRG}=1~~,~~{{\chi_{(2n+1),r}^B}\over
 {\chi_{2,r}^B}}|_{HRG}\simeq \tanh (\mu_B/T)~~ ,~~{{\chi_{2n,r}^B}\over {\chi_{1,r}^B}}|_{HRG}\simeq \coth (\mu_B/T),
\label{ratio}
\end{equation}
which are independent of the number of baryons, their  masses, degeneracy factors or decay widths.

The above structure of cumulants ratios observed  in the HRG  will be modified if the singular part is included in Eq. \ref{obs}.
The  $\chi_{n,s}^{B}$ are increasingly sensitive to the order of  cumulants. Thus,  ratios of cumulants with different $n$  should
 be strongly varying  functions of $T$ and $\mu_B$  when  approaching  a chiral cross over transition.
 Consequently, the observed deviation from that expected in Eq. \ref{ratio} could be considered as a signature of the singular part contribution to the overall
 fluctuations thus, also of the chiral cross over transition.

The generic structure of ratios of different  cumulants  near the chiral cross over transition  is shown in Fig. 1. These ratios were
 calculated in the Polyakov loop extended quark-meson  (PQM)  model at the physical pion mass, applying  the functional renormalization group  (FRG) method.
  In
  FRG  approach  one includes  quantum and thermal fluctuations which are needed to preserve the   universal scaling behavior of physical
   quantities  expected in the O(4) universality class.
In the low temperature phase the ratios  ${{\chi_{2n,r}^B}/ {\chi_{2,r}^B}}=1$ as  expected in the hadron resonance gas from Eq. \ref{ratio}.
  For $n=2$, the kurtosis $R_{4,2}=9{{\chi_{4,r}^B}/ {\chi_{2,r}^B}}$ is not affected  by the chiral
 critical dynamics since there is  no contribution of the singular part to the first four  moments as seen in   Eq. \ref{fluct_mass}.
  The observed in Fig. 1  drop in $R_{4,2}$
  is due to "statistical confinement"
property of the PQM model \cite{stok}. Such behavior of kurtosis was first observed in LQCD calculations and was interpreted as  being a signature
of deconfinement in QCD \cite{s1}.

The large deviations of $(R_{4,2}/9)$ from  unity in  hadronic phase,   which are increasing with $\mu/T$,  are due to a singular contribution   to the
$  {\chi_{4}^B}$ and  $ {\chi_{2}^B}$  ratio (see Eqs. \ref{obs} and \ref{fluct_mass}). With increasing order of cumulants and the value of the chemical potential their  ratios
 are dominated by the   singular part already deeply below  the chiral cross over transition temperature.
   Such behavior could be observed in HIC
 if  freezeout
   appears near the chiral transition.

Recently,  the first data on charge fluctuations and higher
order cumulants, identified through the net-proton fluctuations, were obtained by the  STAR Collaboration in Au-
Au collisions at several collision energies \cite{star}. To explore possible signs of chiral criticality and a cross over transition, the STAR data on the
first four moments  are  compared in Fig. 2   to  HRG  \cite{o1,o3} following Eq. \ref{ratio}. Different ratios of cumulants of the net proton number can be
directly connected with measured
mean (M), variance $(\sigma)$,   skewness (S) and kurtosis $(\kappa)$  \cite{o1}.

 The basic properties of
measured fluctuations and ratios of cumulants are consistent with  that expectated in the  HRG  model \cite{o1,o3}. This  indicates,
that moments of the net proton number
are of thermal origin with respect to  the grand canonical ensemble and that
they  freezeout along the same chemical freezout line as particle yields, close to the chiral cross over line.
However, already such  first comparison of the HRG model with STAR data reveals that
there are deviations \cite{o1,o3,cp1}. This is seen in Fig. 2 on the level of different ratios of cumulants  as well as by comparing directly
 the measured probability distributions
with Skellam distribution expected in the HRG for the net proton number \cite{o3}. The HRG model, as  seen
in Fig. 2, results in a   broader distribution than observed in data,  particularly
at the most central collisions. This implies deviations of data on different moments from the HRG model
which are  increasing with centrality. Shrinking of  measured widths of the  net proton number distribution relative  to the HRG results, seen in Fig. 2,
 is to be already expected  due  to  deconfinement properties of QCD \cite{o3}. We have to stress, however, that the experimental net-proton distributions
  in
 Fig. 2 are not corrected for event-by-event proton/anti-proton  counting efficiency. While the HRG lines in this figure
used efficiency corrected mean multiplicities   from the same experiment.  If  uncorrected
data are used in the Skellam distribution, then the agreement of the HRG model and measure probability distributions
 is found to be much better \cite{cp2}.

From the model calculations in Fig. 1, as well as from Eqs. \ref{obs} and \ref{fluct_mass},
it is clear that contributions of the O(4) singular part to fluctuations
increase
with the order of cummulants.
 Preliminary  data of STAR Collaboration on
$\chi_6^B/\chi_2^B$ ratio, taken in
Au-Au collisions
 at the top RHIC energy,
show a strong deviation of this ratio from the HRG model expectations with a rather  moderate deviations  of the lower order cumulant ratios.
Recently, the STAR Collaboration has also observed  that deviations in central Au-Au collisions  of $\chi_4^B/\chi_2^B$ ratio from the HRG
is non-monotonic in energy.
 Such behaviors of ratios of cumulants  are to be expected due to
remnants of criticality related with a chiral cross over transition.
It is a further  challenge to understand and quantify   the observed energy and centrality dependence  of these  deviations  from  the HRG results.


\vskip 0.1cm
I am grateful for fruitful collaboration and discussions  with P. Braun-Munzinger, B. Briman, F. Karsch and V. Skokov.
Partial support by MEN (Poland) and EMMI (Germany) and discussion with Nu Xu
 is also acknowledged.

\vspace*{-0.5cm}

\end{document}